\documentclass[review]{elsarticle}
\usepackage{hyperref}
\usepackage[english]{babel}
\usepackage{fontenc}
\usepackage{graphicx}
\usepackage{amsmath,amsthm,amssymb}
\usepackage[draft]{changes}
\usepackage{subfigure}
\usepackage{ifthen}
\usepackage{alltt,epstopdf}
\usepackage{enumerate}
\usepackage[text={6.35in,8.55in},centering]{geometry}
\usepackage{tikz}

\newcommand{\pr}{{\bf Proof}~~}

\begin{document}

\begin{frontmatter}

\title{ Modeling and forecasting of the COVID-19 pandemic in India}

\author{Kankan Sarkar}
\address{Department of Mathematics, Malda College, Malda, West Bengal 732101, India}
\fntext[1]{E-mail: subhas.maths@presiuniv.ac.in}

\author{Subhas Khajanchi}
\address{Department of Mathematics, Presidency University, 86/1 College Street,  Kolkata 700073, India}

\date{\today}

\begin{abstract}
\textbf{Background:} The ongoing COVID-19 epidemic dilated rapidly throughout India. To end the global COVID-19 pandemic major behavioral, social distancing, contact tracing, and state interventions has been undertaken to reduce the outbreak and avert the persistence of the coronavirus in humans in India and worldwide. In absence of any vaccine or therapeutics, forecasting is of utmost priority for health care planning and control the transmission of COVID-19.

\textbf{Methods:} We have proposed a mathematical model that explain the transmission dynamics of COVID-19 in India.  Based on the estimated data our model predicts the evolution of epidemics and the end of SARS-CoV-2 and aids to evaluate the influence of different policies to control the spread of the diseases.

\textbf{Findings:} With the real data for infected individuals, we find the basic reproduction number $R_0$ for 17 states of India and overall India. A complete figure is given to demonstrate the estimated pandemic life cycle along with the real data or history to date. Our study reveals that the strict control measures implemented in India substantially mitigated the disseminate of SARS-CoV-2.  Importantly, model simulations predict that 95\% reduction of outbreak on June 26, 2020 and 99\% reduction of outbreak on July 26, 2020 in India.

\textbf{Interpretation:} Our model simulation demonstrates that the community-wide elimination of SARS-CoV-2 is possible by mitigating the social distancing and use essential precautions. Lockdown can be implemented strictly to prevent the human-to-human transmission of COVID-19. The model-based and parameter estimation of epidemic life cycle, and end dates, if can be done precisely, may decrease distress and over optimism and develop the mentality for all of us for the next stages of the outbreak evolution.
\end{abstract}


\end{frontmatter}


\section{Introduction}
\label{Intro}

The ongoing coronavirus, SARS-CoV-2 epidemic has been announced a pandemic by the World Health Organization (WHO) on March 11, 2020 \cite{A-who}, and in the first phase the Govt. of India has announced 21 days nationwide lockdown from March 25, 2020  to April 14, 2020, and in the second phase the lockdown has been extended to May 03, 2020 to prevent stage-III spreading of the virus or human-to-human transmission \cite{Indiacov19}. According to the WHO report dated April 09, 2020 reported 3,855,812 total confirmed cases and 2,65,862 confirmed deaths worldwide \cite{Who-46}. COVID-19 or SARS-CoV-2 has already surpassed the earlier history of two coronavirus outbreaks, namely SARS-CoV (Severe Acute Respiratory Syndrome coronavirus) and MERS-CoV (Middle East respiratory syndrome coronavirus), posing the substantial endanger to the world-wide public health as well as global economy after the 2nd world war \cite{bbc}. SARS-CoV-2 exhibits distinctive epidemiological traits collated with coronavirus epidemics of SARS-CoV and MERS-CoV.

The outbreak was first announced by the ``Health Commission of Hubei province'', China, a cluster of unexplained cases of pneumonia of unknown etiology (unknown cases) \cite{Zhu20}, which is lethal, was first identified in Wuhan city of Hubei province, China, on December 31, 2019 \cite{Wu20,Cohen20}. After that an exceptionally large number of patients were diagnosed with SARS-CoV-2 in mainland of China, prodding Chinese Governments to initiate stringent measures to control the epidemic \cite{Zhu20}. In spite of these precautions, SERS-CoV-2 pandemic evolved in the following months. The confirmed cases of COVID-19, the symptoms like sneeze, fever, or even a runny nose, dry cough, fatigue, breathing problem and bilateral lung infiltration to severely ill and dying.  Due to the human mobility, this communicable disease has now spread throughout the world, making USA and Europe as new epicenters \cite{bbc}.

The first indigenous case of COVID-19 in India was reported on January 30, 2020 in Thrissur district of Kerala and the patient, a student of Wuhan University, China \cite{GoI-1st}. The authorities recommend the level of infection could be assload as the India's testing rates are very poor among the world \cite{bbc1}. The rate of infection of SARS-CoV-2 in India is chronicled to be 1.7 (that is, one coronavirus positive infects 1.7 in India), materially lower than in hot zones \cite{IndExp}. The estimated basic reproduction number $R_0$ for COVID-19 ranges from 2.0 to 3.5 \cite{Anderson20,Zhao20}, that seems similar, or perhaps higher than that of SARS-CoV and MERS-CoV. Higher level of viral loads for COVID-19 were observed in upper respiratory specimens of symptomatic patients resulting little or no symptoms, with a viral shedding pattern similar to that of influenza viruses \cite{Zou20}. Thus, uncertain viral transmission may take part a crucial and underestimated role in sustaining the epidemic.

India has suspended all the tourist visas as the majority of confirmed coronavirus cases were connected to other countries \cite{NdTv}. Governments will not be competent to reduce both fatalities from SARS-CoV-2 epidemic and the economic impact of viral outbreak. Maintaining the fatality rate as low as possible will be the utmost importance for the populations; therefore the Governments must put in place measures to mitigate the unavoidable economic downturn. Due to absence of any specific pharmaceutical interventions, government of various countries are imposing different strategies to prevent this outbreak and the lockdown is the most common one. As for examples, the measures adopted in this time incorporated social distancing, closing schools, universities, offices, churches, bars, avoid mass gatherings,  other social places as well as contact of cases (quarantine, surveillance, contact tracing) \cite{Ferguson20}. On March 19, 2020 the Govt. of India suspended all the international flights till March 22, 2020 \cite{Indiatvnews}, and on March 23, 2020 the union Govt. also suspended all the domestic flights till March 25, 2020 \cite{Livemint1} to maintain the social distancing among the people. The prime minister of India has announced a 14 hours voluntary public curfew ('Janata Curfew') on March 22, 2020 as a precautionary measure to combat against COVID-19. The Govt. of India followed it up with lockdowns on March 23, 2020 to prevent the emanating threat in 75 districts across the country including major cities where the COVID-19 infection was endemic \cite{ReganCNN}. Furthermore, on March 24, 2020 the Govt. of India has ordered a nationwide lockdown for 21 days, overwhelming the entire 1.3 billion public in India \cite{NdTv1}, and the lockdown has been extended to May 03, 2020 to prevent stage-III spreading of the virus or human-to-human transmission \cite{Livemint}.

\begin{figure}[tbh!]
  \centering
  \includegraphics[width=1.0\textwidth]{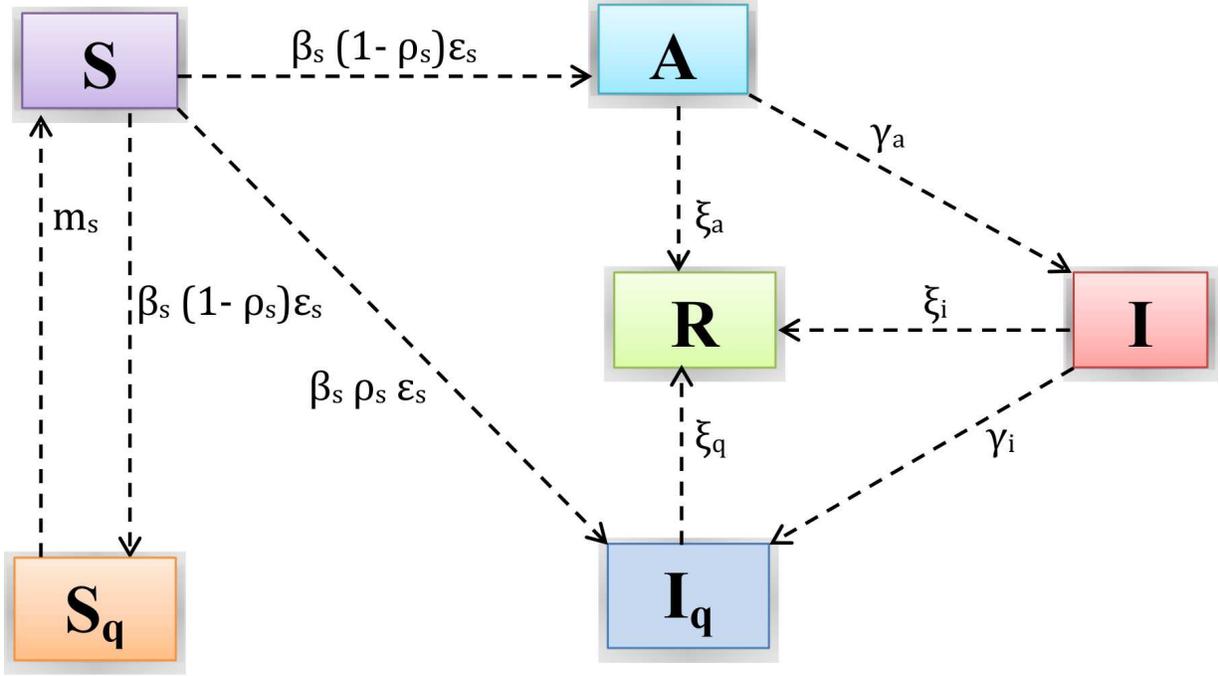}
  \caption{Schematic representation of the model. The schematic flow diagram represents the interplays among different stages of infection in the model $SARII_qS_q$: susceptible or uninfected $(S)$, asymptotic or mildly symptomatic $(A)$, recovered or healed $(R)$, infected or symptomatic $(I)$, isolated infected $(I_q)$ and quarantined susceptible $(S_q)$ individuals. }
\label{Schema}
\end{figure}

Predictive mathematical models play a key role to understand the course of the epidemic and for designing strategies to contain quickly spreading infectious diseases in lack of any specific antivirals or effective vaccine \cite{Anderson91,Diekmann00,Hethcote00}. In the year 1927, Kermack \& McKendrick \cite{Kermack27} developed a fundamental epidemic model for human-to-human transmission to describe the dynamics of populations through three mutually exclusive phages of infection, namely susceptible $(S)$, infected $(I)$ and removed $(R)$ classes. Mathematical modeling of infectious diseases are now ubiquitous and many of them can precisely depict the dynamic spread of particular epidemics. Several mathematical models has been developed to study the transmission dynamics of COVID-19 pandemic. A Bats-Hosts-Reservoir-People network model has been developed by Chen et al. \cite{Chen20a} to study the transmission dynamics of novel coronavirus. Lin et al. \cite{Lin20} extended the SEIR (susceptible-exposed-infectious-removed) compartment model to study the dynamics of COVID-19 incorporating public perception of risk and the number of cumulative cases. Khajanchi et al. \cite{Kh1Cov20} studied an extended SEIR model to study the transmission dynamics of COVID-19 and perform a short-term prediction based on the data from India. A discrete-time SIR (susceptible-infectious-removed) model introducing dead compartment system studied by Anastassopoulou et al. \cite{Anastassopoulou} to portray the dynamics of COVID-19 outbreak. Wu et al. \cite{Wu20} studied a SEIR model to investigate the dynamics of 2019-nCoV human-human transmission dynamics based on the data from Wuhan, China from December 31, 2019 to January 28, 2020 and calculate the basic reproduction number was approximately 2.68. Wu et al. \cite{Wu20a} used a SIR model to delineate the transmission dynamics of COVID-19 and also estimate the clinical severity for the coronavirus. To study the dynamics of COVID-19, a stochastic transmission model also developed by Kucharski et al. \cite{Kucharski20}. Here, we developed a new epidemiological mathematical model for novel coronavirus or SARS-CoV-2 epidemic in India that extends the standard SEIR compartment  model, alike to that studied by Tang et. al. \cite{Tang20} for COVID-19. The transmission dynamics of our proposed model for COVID-19 is illustrated in the Figure \ref{Schema}.

\section{Dynamic model without effective control measures}
\label{model}

We develop here a classical SEIR (susceptible-exposed-infectious-recovered)-type epidemiological model by introducing contact tracing and other interventions such as quarantine, lockdown, social distancing and isolation that can represent the overall dynamics of novel coronavirus or COVID-19 (SARS-CoV-2). The model, named $SARII_{q}S_{q}$, monitors the dynamics of six compartments (classes), namely susceptible individuals  $(S)$ (uninfected), quarantined susceptible individuals $(S_q)$ (quarantined at home), infectious but not yet symptomatic or asymptomatic infectious individuals $(A)$, infected or infectious with symptoms/clinically ill $(I)$, isolated infected individuals $(I_q)$ (infected or life-threatening or detected) and the recovered compartment $(R)$ (no more infectious). The total size of the individuals is $N = S + S_q + A + I + I_q + R$. In our model, quarantine describes the separation of coronavirus infected populations from the susceptible individuals before progression of clinical symptoms, whereas the isolation refers to the dissociation of coronavirus infected populations with such clinical symptoms. The rate of change in each compartments at any time $t$ is represented by the following system of nonlinear ordinary differential equations:

\begin{eqnarray}
\label{stateeq}
\left\{
  \begin{array}{ll}
    \frac{dS}{dt} = \Lambda_{s} - (\beta_s + \rho_s (1 - \beta_s)) \varepsilon_{s} S\frac{I}{N} - \delta S + m_{s} S_{q}, \\[0.10cm]
   \frac{dS_q}{dt} = (1 - \beta_s) \varepsilon_{s} \rho_s S \frac{I}{N} - (m_s + \delta) S_{q}, \\[0.10cm]
   \frac{dA}{dt} = \beta_s (1 - \rho_s) \varepsilon_s S\frac{I }{N} - (\gamma_a + \xi_a + \delta)A,  \\[0.10cm]
    \frac{dI}{dt} = \gamma_a A - (\gamma_i + \xi_i + \delta)I, \\[0.10cm]
    \frac{dI_{q}}{dt} = \beta_{s} \varepsilon_s \rho_s S \frac{I}{N} + \gamma_i  I - (\xi_q + \delta) I_q, \\[0.10cm]
    \frac{dR}{dt} = \xi_a A + \xi_{i}I + \xi_{q} I_q - \delta R,
     \end{array}
\right.
\end{eqnarray}
the model is supplemented by the following non-negative initial values:
\begin{eqnarray}
\label{IC}
S(t_0) = S_0,~~~S_{q}(t_0) = S_{q0},~~~A(t_0) = A_0,~~~I(t_0) = I_0,~~~I_q(t_0) = I_{q0}~~\textrm{and}~~R(t_0)=R_0.
\end{eqnarray}
Herein, $t \geq t_0$ represents time in days and $t_0$ indicates the starting date for the system of the coronavirus epidemic.

In our model construction, $\beta_s$ represents the probability of transmission per contact between an infective and a susceptible class, and  $\varepsilon_s$ is denoted by the daily contact rate per unit of time. Here the parameter $\beta = \beta_s \varepsilon_s$ is explicitly associated with the measures like lock-down, social distancing, shaking hand, coughing and sneezing etc., which exactly decrease the number of social contacts. By enforcing contact tracing, a proportion $\rho_s$, of individuals exposed to the coronavirus is quarantined. The quarantined classes can either move to the compartment $S_q$ or $I_q$, depending on whether they are effectively infected individuals or not, whereas the another proportion $1-\rho_s$, consists of populations exposed to the coronavirus who are missed from contact tracing and move to the infectious class $I$ (once infected) or remaining in susceptible class $S$ (if uninfected). Then the quarantined classes, if uninfected (or infected), move to the class $S_q$ (or $I_q$) at a rate of $(1 - \beta_s) \rho_s \varepsilon_s$ (or $\beta_s \rho_s \varepsilon_s $). Those who are not quarantined individuals, but asymptomatic infectious individuals, will move to the asymptomatic compartment $A$ at the rate of $\beta_s (1 - \rho_s) \varepsilon_s$. The proportion $\rho_s$ ($0 < \rho_s < 1$) encapsulates the effectiveness of different individual preventive measures, such as cough etiquettes, frequently wash hand by soap and use sanitizer etc. We utilize a constant $m_s$ to indicate the transition rate from the susceptible individuals to the (COVID-19) quarantined susceptible class due to fever and/or illness-like clinical symptoms. We symbolize $\xi_a$, $\xi_i$ and $\xi_q$ are the rates of recovery individuals of asymptomatic class, symptomatic or clinically ill patients and isolated individuals, respectively. Our model introduces some demographic effects by considering a proportional natural decay rate $\delta$ in each of the six individuals, and $\Lambda_s$ represents the constant inflow of susceptible individuals. Asymptomatic population develop to infected population at the rate $\gamma_a$, so the average time spent in the asymptotic class is $\frac{1}{\gamma_a}$ per unit time. In similar fashion, $\frac{1}{\gamma_i}$ represents the mean duration for infected individuals. We ignore the rate of probability of transforming susceptible again after having cured (recovered) from the disease infection. It is to be noted that our $SARII_qS_q$ model did not take into account many important ingredients that take part a key role in the transmission dynamics of COVID-19 such as the influence of the latency period, the inhomogeneous disease transmission network, the influence of the measures already considered to fight the coronavirus diseases, the features of the individuals (for example, the influence of the stage-structure, individuals who are already medically unfit).

\subsection{Basic reproduction number}
The basic reproduction number, symbolized by $R_0$, is `the expected number of secondary cases produced, in a completely susceptible population, by a typical infective individual' \cite{Diekmann90,PVDD02}. The dimensionless basic reproduction number provides a threshold, which play a crucial role in determining the disease persists or dies out from the individual. In a more general way the basic reproduction number $R_0$ can be stated as the number of new infections created by a typical infective population at a disease free equilibrium. $R_0 < 1$ determines on average an infected population creates less than one new infected population during the course of its infective period, and the infection can die out. In reverse way, $R_0 > 1$ determines each infected population creates, on average, more than one new infection, and the disease can spread over the population. The basic reproduction number $R_0$ can be computed by using the concept of next generation matrix \cite{Diekmann90,PVDD02}. In order to do this, we consider the nonnegative matrix $\mathcal{F}$ and the non-singular $M-$matrix $\mathcal{V}$, expressing as the production of new-infection and transition part respectively, for the system (\ref{stateeq}), are described by
\begin{eqnarray*}
\mathcal{F} &=& \left[ \begin{array}{c}
      \beta_s(1 - \rho_s)\varepsilon_s S \frac{I}{N}  \\
       0 \\
       \beta_s \varepsilon_s \rho_s S \frac{I}{N}
     \end{array} \right],~~~~~~~~~\mathcal{V} = \left[ \begin{array}{c}
       (\gamma_a + \xi_a )A  \\
       - \gamma_a A + (\gamma_i + \xi_i)I \\
       -\gamma_i I + \xi_q I_q
     \end{array} \right].
\end{eqnarray*}

The variational matrix of the model (\ref{stateeq}) computed at the infection free state $(S_{q} = A = I = I_{q} = R = 0)$ gives
\begin{eqnarray*}
F &=& \left[ \begin{array}{ccc}
       0 & \beta_s(1 - \rho_s)\varepsilon_s \frac{S}{N} & 0  \\
       0 & 0 & 0  \\
       0 & \beta_s \varepsilon_s \rho_s \frac{S}{N} & 0
     \end{array} \right],
~~~~~~~~V ~=~ \left[ \begin{array}{ccc}
       \gamma_a + \xi_a & 0 & 0  \\
       - \gamma_a & \gamma_i + \xi_i & 0  \\
       0 & - \gamma_i & \xi_q
     \end{array} \right].
\end{eqnarray*}

The basic reproduction number $\mathcal{R}_0 = \rho(FV^{-1})$, where $\rho(FV^{-1})$ represents the spectral radius for a next generation matrix $FV^{-1}$. Thus, the basic reproduction number of the system (\ref{stateeq}) is
\begin{eqnarray}
\label{R0}
\rho(F V^{-1}) &=& \mathcal{R}_{0} ~=~ \frac{\beta_s (1 - \rho_s)\varepsilon_s \gamma_a}{ (\gamma_a + \xi_a) (\gamma_i + \xi_i)}.
\end{eqnarray}

\begin{table}[ht]
  \caption{Parameters of the $SARII_qS_q$ model system (\ref{stateeq}) and their biological interpretations.}
\label{parval}
 {\begin{tabular}{p{0.8cm} l l l}
  \\[-0.1cm]
  \hline
  Symbol        & Description & Values \& Source   \\[0.01cm]
  \hline
 $\Lambda_s$    & net inflow rate of susceptible individuals & 0 \\
  $\beta_s$     & probability of disease transmission & Estimated \\
  $\rho_s$      & quarantined rate of susceptible individuals & Estimated \\
$\varepsilon_s$ & contact rate of entire individuals & Estimated \\
$\delta  $      & natural mortality rate & 0 \\
$m_s$           & rate at which quarantined susceptible individuals are released into uninfected class & $\frac{1}{24}$ \cite{Tang12} \\
$\gamma_a$      & probability rate at which asymptomatic individuals develops clinically symptoms &  Estimated \\
$\xi_a$         & recovery rate of asymptomatic infected individuals & Estimated \\
$\gamma_i$      & probability rate at which infected individuals become isolated & $7.151 \times 10^{-2}$ \cite{Tang20} \\
$\xi_i$         & rate of recovery for infected individuals & Estimated \\
$\xi_q$         & recovery rate of isolated infected individuals & $\frac{1}{7.48}$ \cite{Tang12}  \\
  \hline
  \end{tabular}}
\end{table}

\section{Model calibration and data source}

We calibrated our $SARII_qS_q$ model for COVID-19 to the daily new infected cases and cumulative confirmed cases of SARS-CoV-2 viruses for 17 provinces of India, namely Andhra Pradesh, Delhi, Gujarat, Hariyana, Jammu \& Kashmir, Karnataka, Kerala, Madhya Pradesh, Maharashtra, Punjab, Rajasthan, Tamil Nadu, Telangana, Uttar Pradesh, West Bengal, Bihar, Odisha and the overall India with real data till April 30, 2020. The description of the $SARII_qS_q$ model are given in Table \ref{parval}, list of key estimated parameter values are specified in Table \ref{parvalEst} and estimated initial population size are given in the Table \ref{parvalIC}. By calibrating the $SARII_qS_q$ model parameters with real data up to 30 April 2020, we make an attempt to forecast the evolution of the epidemic in India and 17 provinces of India. In the model exploration, we did not consider the demographic effects because of the short epidemic time scale in compare to the demographic time scale, that is, $\Lambda_s = \delta = 0$.

\subsection{Parameter estimation}

The most important challenge in any mathematical model based study is to estimate the model parameters and the initial population size.  The solution of the $SARII_qS_q$ model system (\ref{stateeq}) depends on both the parameter values and initial population size. The model parameters have been estimated assuming the initial population size and fitting the model simulation with the observed COVID-19 cases.  The assumed initial population sizes are presented in the Table \ref{parvalIC}. We have estimated six parameters, probability of disease transmission $(\beta_s)$, quarantined rate of susceptible individuals $(\rho_s)$, contact rate of entire individuals $(\epsilon_s)$, probability rate at which asymptomatic individuals develop clinical symptoms $(\gamma_a)$, recovery rate of asymptomatic infected individuals $(\xi_a)$ and rate of recovery for infected individuals $(\xi_i)$. The parameters are estimated from the observed daily new COVID-19 or SARS-CoV-2 viruses.  Although, we have shown the plot validating model simulation for cumulative cases, the parameters are not estimated from confirmed cumulative COVID-19 cases to optimize the error in parameter estimation \cite{King15} and errors are listed in the Table \ref{Table:error}.

\section{Numerical simulation}

\begin{figure}
\centering
\includegraphics[width=17cm]{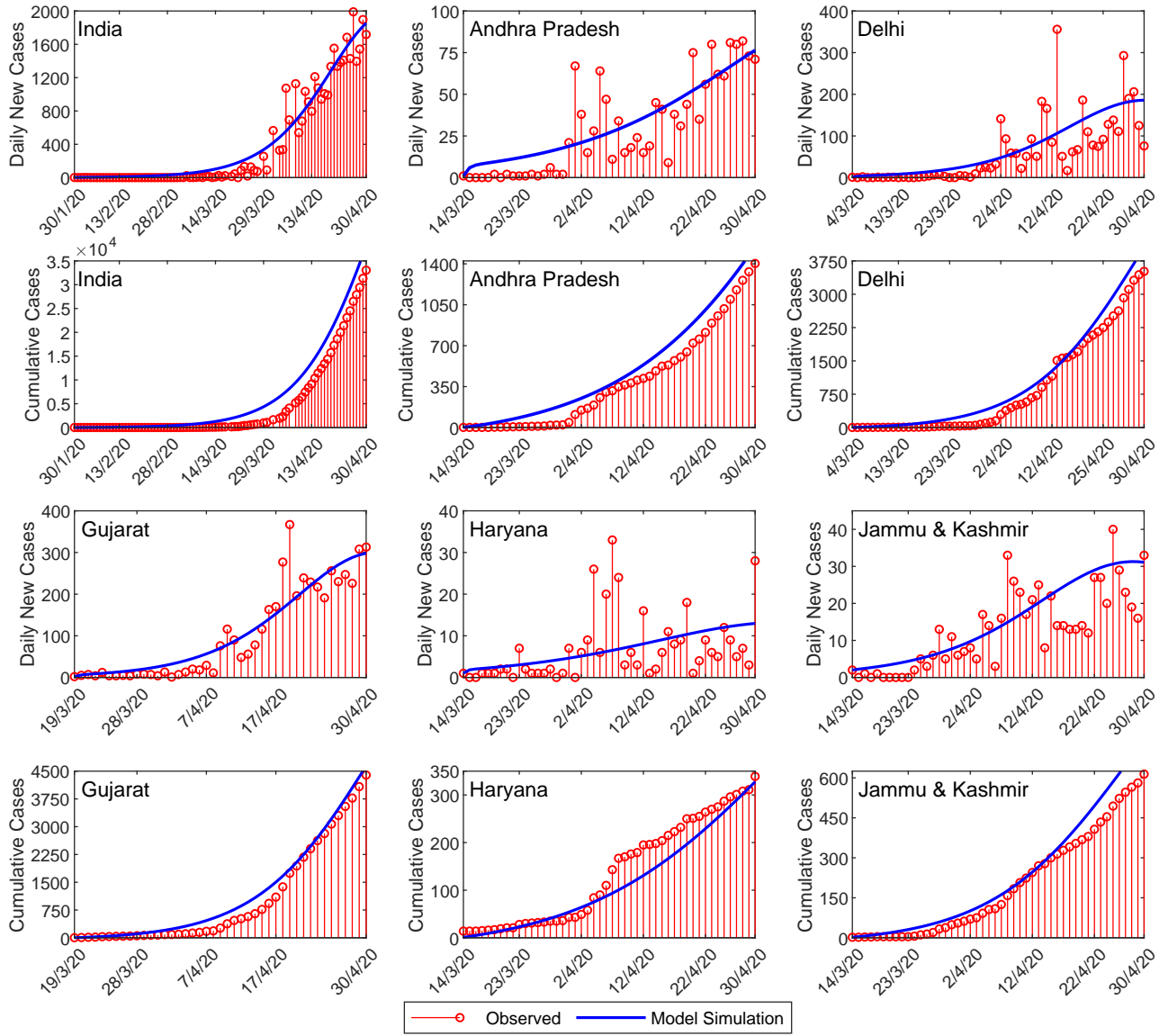}
\caption{Model estimation based on real data. The $SARII_qS_q$ model fitted with the real data on daily new cases of COVID-19 in India and 5 provinces of India, namely Andhra Pradesh, Delhi, Gujarat, Haryana and Jammu \& Kashmir. Observed data points are displayed in the red dot histogram and the blue curve represents the best fitting curve for the $SARII_qS_q$ model. The first and third rows represents the daily new cases of coronavirus diseases, whereas the second and fourth rows represents the cumulative confirmed cases of COVID-19. The estimated parameter values are listed in the Table \ref{parvalEst}. The initial values used for this parameter values are presented in the Table \ref{parvalIC}.}
\label{f_DataFit1}
\end{figure}

\begin{figure}
\centering
\includegraphics[width=17cm]{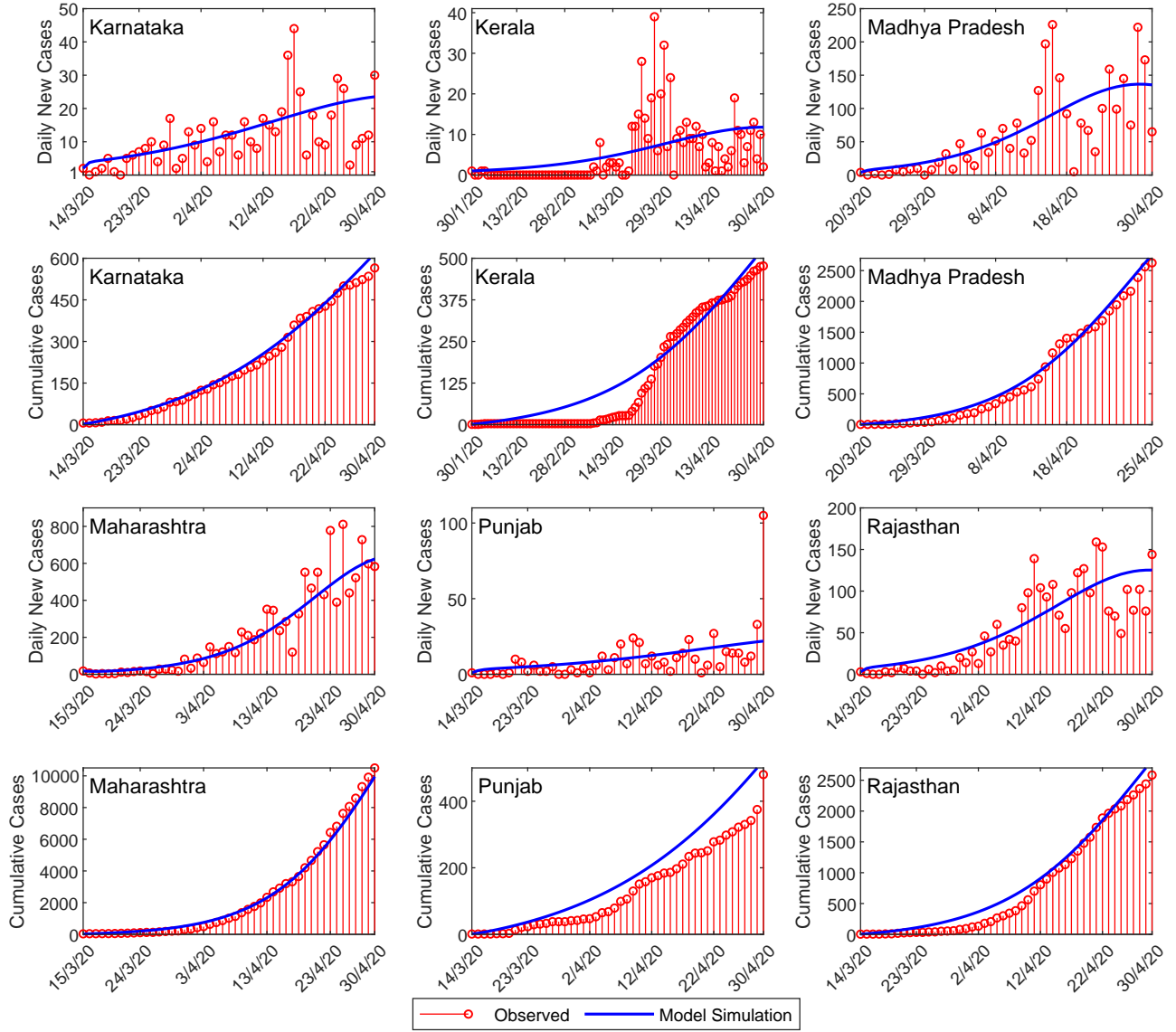}
\caption{Model estimation based on real data.  The $SARII_qS_q$ model fitted with the real data on daily new cases of COVID-19 for 6 provinces of India, namely Karnataka, Kerala, Madhya Pradesh, Maharashtra, Punjab, and Rajasthan. Observed data points are shown in the red dot histogram and the blue curve represents the best fitting curve for the $SARII_qS_q$ model. The first and third rows represents the daily new cases of coronavirus diseases, whereas the second and fourth rows represents the cumulative confirmed cases of COVID-19. The estimated parameter values are listed in the Table \ref{parvalEst}. The initial values used for this parameter values are presented in the Table \ref{parvalIC}. }
\label{f_DataFit2}
\end{figure}

\begin{figure}
\centering
\includegraphics[width=17cm]{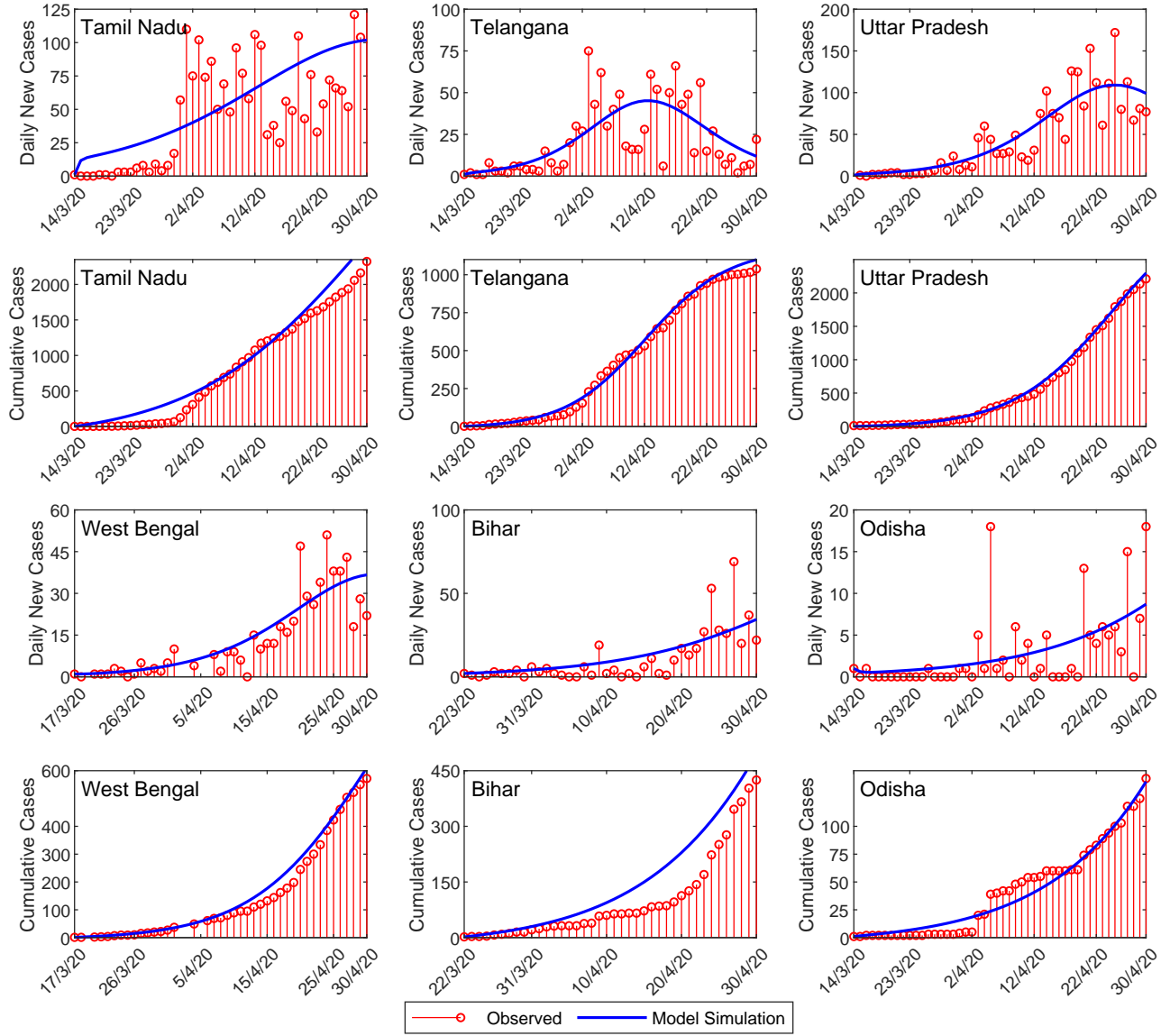}
\caption{Model estimation based on real data.  The $SARII_qS_q$ model fitted with the real data on daily new cases of COVID-19 for 6 provinces of India, namely Tamil Nadu, Telangana, Uttar Pradesh, West Bengal, Bihar, and Odisha. Observed data points are shown in the red dot histogram and the blue curve represents the best fitting curve for the $SARII_qS_q$ model. The first and third rows represents the daily new cases of coronavirus diseases, whereas the second and fourth rows represents the cumulative confirmed cases of COVID-19. The estimated parameter values are listed in the Table \ref{parvalEst}. The initial values used for this parameter values are presented in the Table \ref{parvalIC}. }
\label{f_DataFit3}
\end{figure}

Initially we have validated the model simulation with the observed COVID-19 cases. Model simulated from the first date of coronavirus infection and up to 30 April, 2020 for whole India and for seventeen states of India. The model simulation fitted with the observed daily new COVID-19 cases and cumulative COVID-19 cases. The parameter values are taken from Table \ref{parval} and the Table \ref{parvalEst} and the initial population size from Table \ref{parvalIC}. Figures \ref{f_DataFit1}, \ref{f_DataFit2} and \ref{f_DataFit3} shows the observed COVID-19 cases and model simulations. Observed data points are displayed in the red dot histogram and the blue curve represents the best fitting curve for the model simulation. The first and third rows in each of these figures represents the daily new cases of coronavirus diseases, whereas the second and fourth rows represents the cumulative confirmed cases of COVID-19. In all the states of India except Telangana, the number of daily new COVID-19 cases is increasing and the model capture this increasing trend very well. However, in Telanga, the number of daily new COVID-19 cases starts decreasing from 12 April, 2020 onwards and our model captures this trend successfully (see the Figure \ref{f_DataFit3}).

To measure the accuracy of fitting, we have computed Mean Absolute Error $(E_{MAE})$ and Root Mean Square Error $(E_{RMSE})$. The $E_{MAE}$ and $E_{RMSE}$ are defined as follows:

\begin{eqnarray}
\begin{aligned}
E_{MAE}=\frac{\Sigma_{i=1}^{n} \vert C(i)-S(i) \vert}{n}, \\[0.15cm]
E_{RMSE}=\sqrt{\frac{\Sigma_{i=1}^{n} ( C(i)-S(i))^2}{n}}
\nonumber
\end{aligned}
\end{eqnarray}

where $C(i)$ represents the observed value, $S(i)$ is the model simulation and $n$ is the sample size of the observed data. The values of $E_{MAE}$ and $E_{RMSE}$ for the seventeen states and for whole India are listed in the Table \ref{Table:error}. Our model performs excellently well in all the provinces. The values of $E_{MAE}$ and $E_{RMSE}$ for whole India are 114.0705 and 187.4646 respectively. Also for all the states of India, the value of $E_{MAE}$ varies from 2.3084 to 53.2686 and the value of $E_{RMSE}$ varies from 3.7892 to 54.9932.

\subsection{Sensitivity analysis for $R_0$}

\begin{figure}
\centering
\includegraphics[width=15cm]{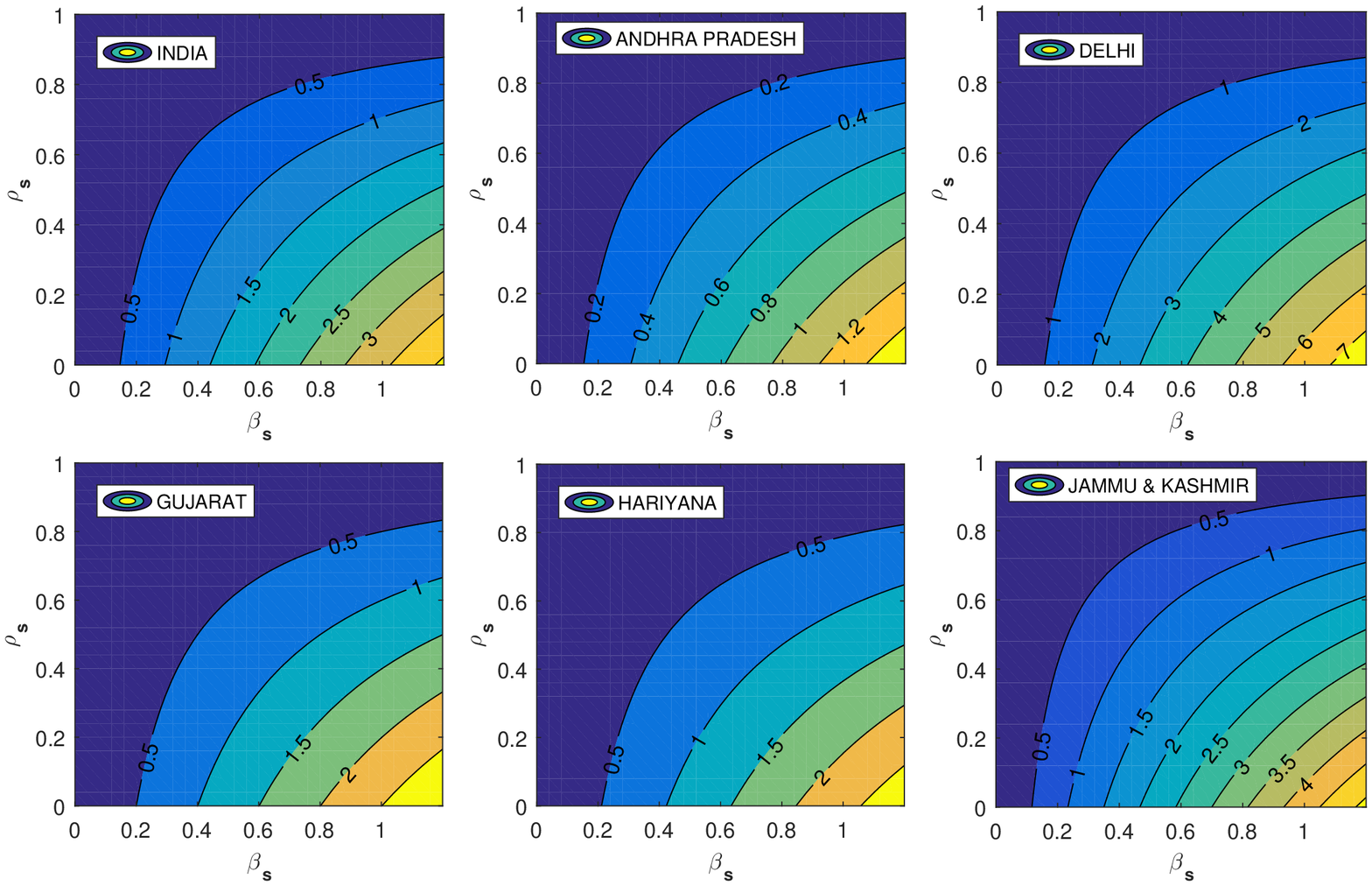}
\caption{Contour plots of basic reproduction number $R_0$. Contour plots of $R_0$ for India and five different provinces of India, with respect to the probability of disease transmission rate $\beta_s$ and the quarantined rate $\rho_s$ of susceptible individuals. All parameter values other than $\beta_s$ and $\rho_s$ are listed in the Table \ref{parvalEst}. The contour plots demonstrates that higher disease transmission probability of the COVID-19 virus will remarkably increase the reproduction number $R_0$. }
\label{f_cp1}
\end{figure}

To describe how best to minimize individuals impermanence and morbidity due to SARS-CoV-2, it is important to see the relative significance of various ingredients responsible for disease transmission. Transmission of SARS-CoV-2 is directly related to the basic reproduction number $R_0$. We compute the sensitivity indices for $R_0$ for the parameters of the $SARII_qS_q$ model. This indices apprise us how important each parameter is to disease transmission. Sensitivity analysis is mainly used to describe the robustness of the model predictions to the parameters, as there are generally errors in collection of data and assumed parameter values. Sensitivity indices quantify the relative change in a state variable when a parameter alters. The normalized forward sensitivity index for $R_0$, with respect to the disease transmission coefficient $\beta_s$ can be defined as follows:
\begin{eqnarray*}
\Upsilon_{\beta_s}^{R_0} &=& \frac{\partial R_0}{\partial \beta_s} \times \frac{\beta_s}{R_0} ~=~ 1,
\end{eqnarray*}
which demonstrates that $R_0$ is a increasing function of $\beta_s$. This implies that probability of disease transmission has a high influence on COVID-19 control and management. The sensitivity indices of other parameters are given in the Table \ref{parvalSI}. In the Table \ref{parvalSI}, some of the indices are positive (and some are negative) which means if the parameter increases then increase the value of $R_0$ (and if the parameter increases then decrease the value of $R_0$). To control the outbreak of SARS-CoV-2, we must select the most sensitive parameters who have most influence to reduce the diseases. As for example, the transmission rate $\beta_s$ has an impact in reducing the COVID-19 diseases, which can easily be observed from the Table \ref{parvalSI}. Therefore, we draw the contour plots for $R_0$ in the Figure \ref{f_cp1} dependence on the rate of disease transmission probability $\beta_s$ and the quarantine rate $\rho_s$. Contour plot shows that for the higher values of $\beta_s$ the reproduction number $R_0$ increases significantly, which means that the SARS-CoV-2 disease will persist among the human and spread throughout the community if the public not take the preventive measures. Thus, to control $R_0$ must reduce the disease transmission coefficient $\beta_s$ and increase the period of quarantine rate $\rho_s$. Thus, we may conclude that to end the COVID-19 outbreak enhance the quarantine and reduce the probability of disease transmission following contact tracing, social distancing, limit or stop theaters and cultural programme etc.

\subsection{Effect of lockdown}

To investigate the influence of intervention policies in reducing the SARS-CoV-2 viruses, we modified the disease transmission rate $\beta_s$ by $(1-\omega)\beta_s$ with $0 < \omega < 1$, where $\omega$ encapsulates the efficacy of different individuals precautionary measures. Intervention strategies including the most recent lockdown progressively implemented since 25 March 2020, have influenced the spread of the outbreak.  We also predict our model for 17 provinces of India and overall India, involving social distancing to control COVID-19.

\begin{figure}
\centering
\includegraphics[width=17cm]{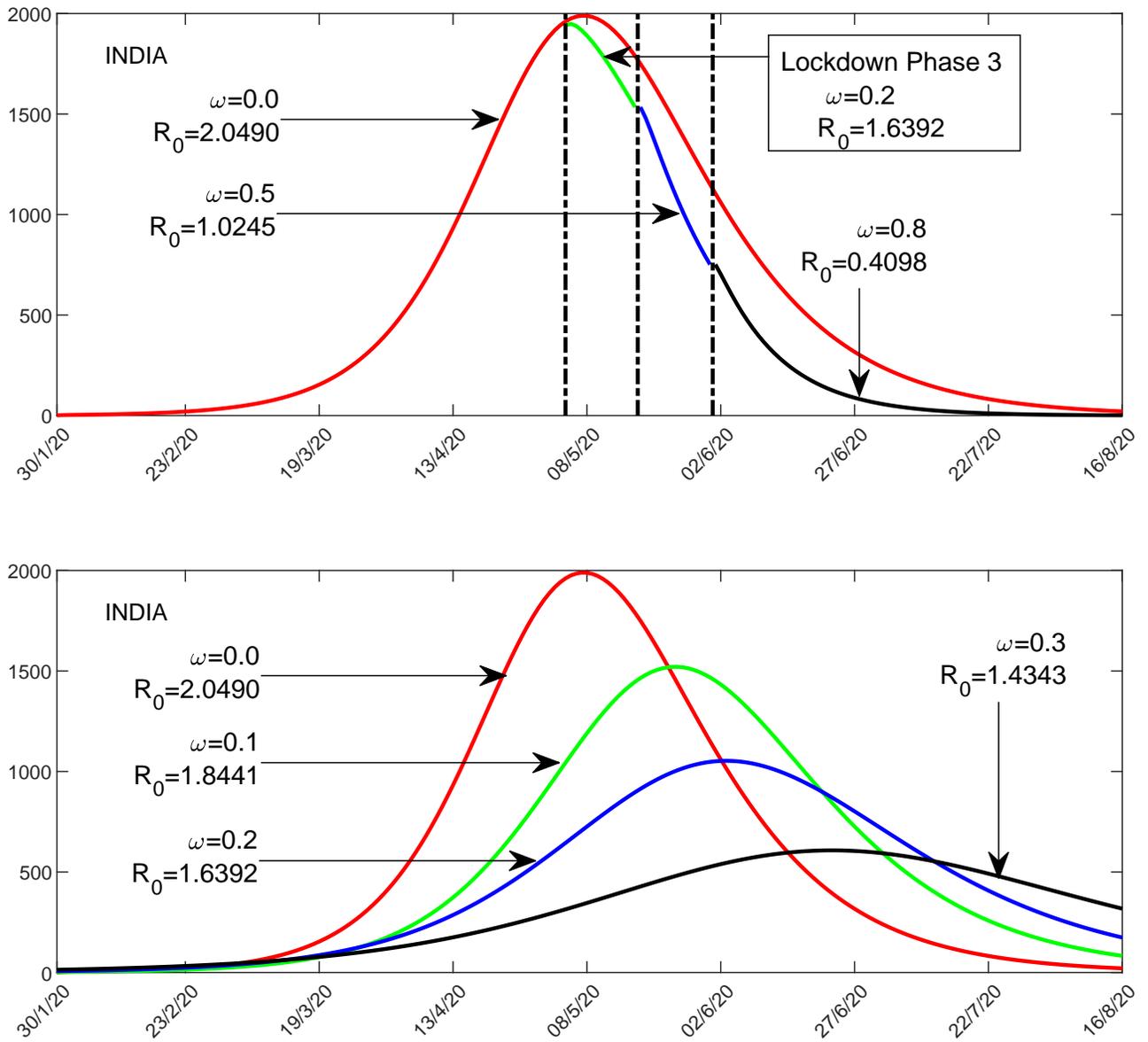}
\caption{Effect of nation-wide lockdown. Epidemic evolution predicted by the $SARII_qS_q$ model for SARS-CoV-2 epidemic in India for the time period 30 January 2020 to 16 August 2020. Before implemented 3rd Phase lockdown, the social distancing are not enforced, resulting a larger $R_0 = 2.0490$. After implementation of 3rd phase lockdown, the social distancing are enforced but not strong enough, as a result larger $R_0 = 1.6302$  and $R_0 = 1.0245$, which shows a substantial outbreak of COVID-19. But after implementation of fully operational and strict lockdown, after 31 May 2020, resulting a smaller $R_0 = 0.4098$, yields below 1. Bottom panel shows the trajectories of COVID-19 diseases for different values of the strengthen of the intervention $\omega$. }
\label{f_lcd}
\end{figure}

For the estimated model parameters, before 3rd phase of lockdown $(\omega = 0)$ the basic reproduction number was $R_0 = 2.0490$ that followed in a significant outbreak of COVID-19. The 3rd phase lockdown in India, has been implemented from 04 May 2020 and we divide into three time periods, first time period from 04 May 2020 to 17 May 2020, second time period from 18 May 2020 to 31 May 2020 and third time period from 01 June 2020 to 16 August 2020.  In the first time window, $R_0 = 1.6392$ as an outcome of the inclusion of lockdown, social distancing, hygiene precautions, and early measures by Indian Govt. (e.g. closing school, shopping mall etc.). In the second time window, the country-wide lockdown was implemented but not yet enforced, as result $R_0 = 1.0245$. The effectiveness of COVID-19 has been reduced due to lockdown but need to impose more restriction. The nation-wide lockdown was implemented in the third time period with strictly enforced, as a result $R_0 = 0.4098$ reached below 1. The top panel of the Figure \ref{f_lcd} shows the effect of lockdown in three different time periods. Figure \ref{f_lcd} predict that on 17 May 2020, the cumulative number of COVID-19 infected population in India without Phase-3 lockdown is 70948 but after strictly enforcing nation-wide Phase-3 lockdown in India the cumulative number of infected population reduced and it will be 64838. Another interesting outcome can be noticed as an effect of nation-wide lockdown is the appearance of epidemic peak. The effect of lockdown not only reduce the infected population but also causes delay in an appearance of the peak, which has shown the bottom panel of the Figure \ref{f_lcd}. The bottom panel of Figure \ref{f_lcd} exhibits that the infected population reaches its peak at a different time.

\subsection{Model prediction}

The first positive COVID-19 cases was observed in India on 30 January, 2020 and 02 March, 2020 onward and multiple new cases have been reported from all over the country. Forecast of the pandemic is most essential to take necessary administrative actions and health care planning and precautions. To estimate the end date of the pandemic, we have numerically simulated $SARII_qS_q$ model and estimated the end days of COVID-19 infection. The numerical simulation of the model depends on the value of the parameters estimated and the parameters have been estimated from the observed data up to 30 April, 2020. So, if more observed data are available, the parameters values will alter and the long time prediction will be more accurate. However, this prediction gives us an overview of the pandemic, which will lead to decide future planning. In this study, we fitted $SARII_qS_q$ model to forecast the pandemic trend over the period after 30 April, 2020 by using the observed data from the first day of infection to 30 April, 2020 at national and state level.

\begin{figure}
\centering
\includegraphics[width=17cm]{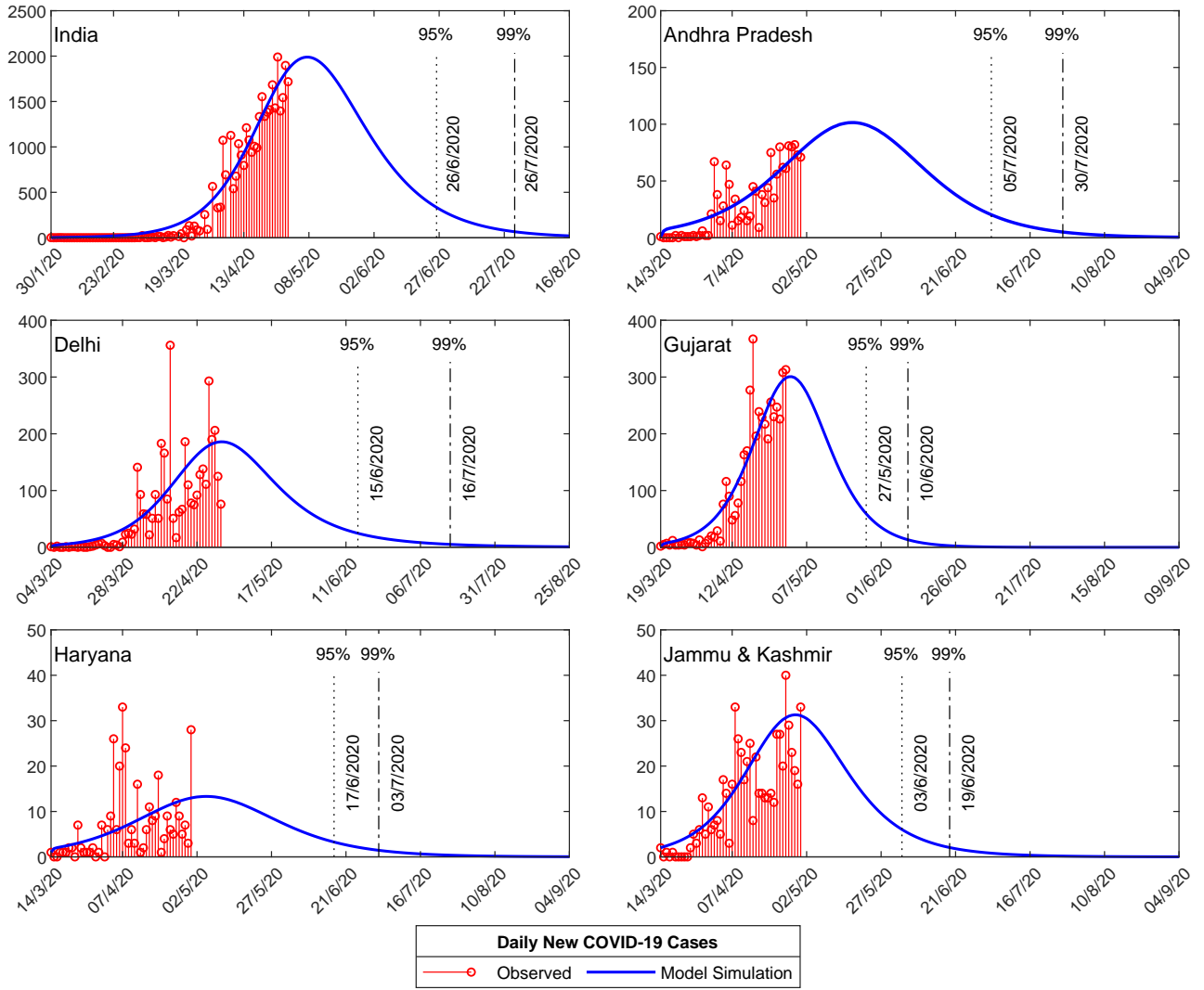}
\caption{Model-Based Data-Driven prediction of SARS-CoV-2 life cycle. Model predictions based on the estimated data for the Republic of India and five provinces of India, namely Andhra Pradesh, Delhi, Gujarat, Haryana and Jammu \& Kashmir. The blue curve represents the model prediction for the daily new confirmed cases of coronavirus diseases while the red dot histogram represents the actual cases. The turning and ending dates of COVID-19 in India and five provinces are displayed in the figure. The estimated parameter values are specified in the Table \ref{parvalEst}. }
\label{f_Predict1}
\end{figure}

\begin{figure}
\centering
\includegraphics[width=17cm]{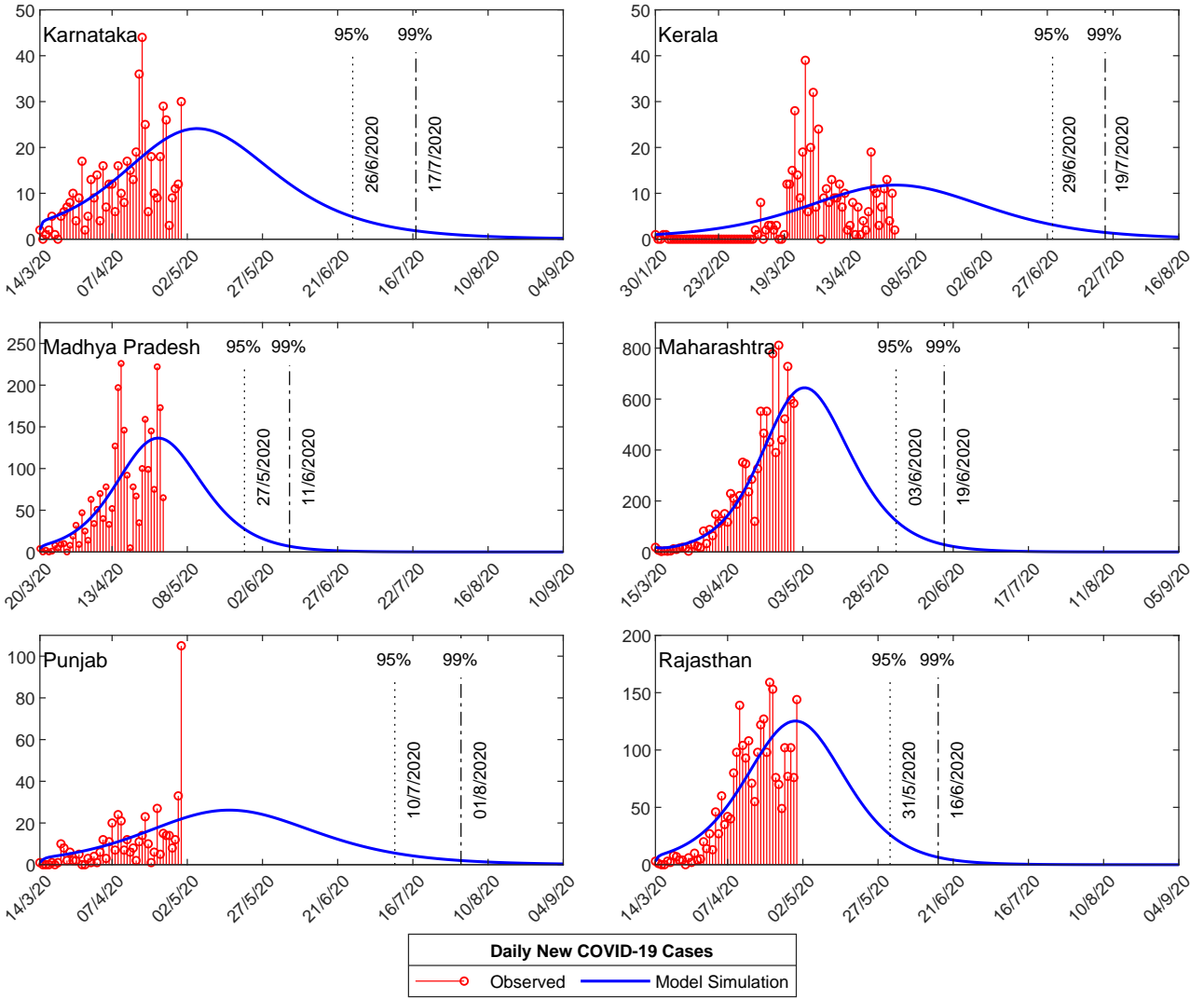}
\caption{Model-Based Data-Driven prediction of SARS-CoV-2 life cycle. Model predictions based on the estimated data for the five provinces of India, namely Karnataka, Kerala, Madhya Pradesh, Maharashtra, Punjab, and Rajasthan. The blue curve represents the model prediction for the daily new confirmed cases of coronavirus diseases while the red dot histogram represents the actual cases. The turning and ending dates of COVID-19 in India and five provinces are displayed in the figure. The estimated parameter values are specified in the Table \ref{parvalEst}.}
\label{f_Predict2}
\end{figure}

\begin{figure}
\centering
\includegraphics[width=17cm]{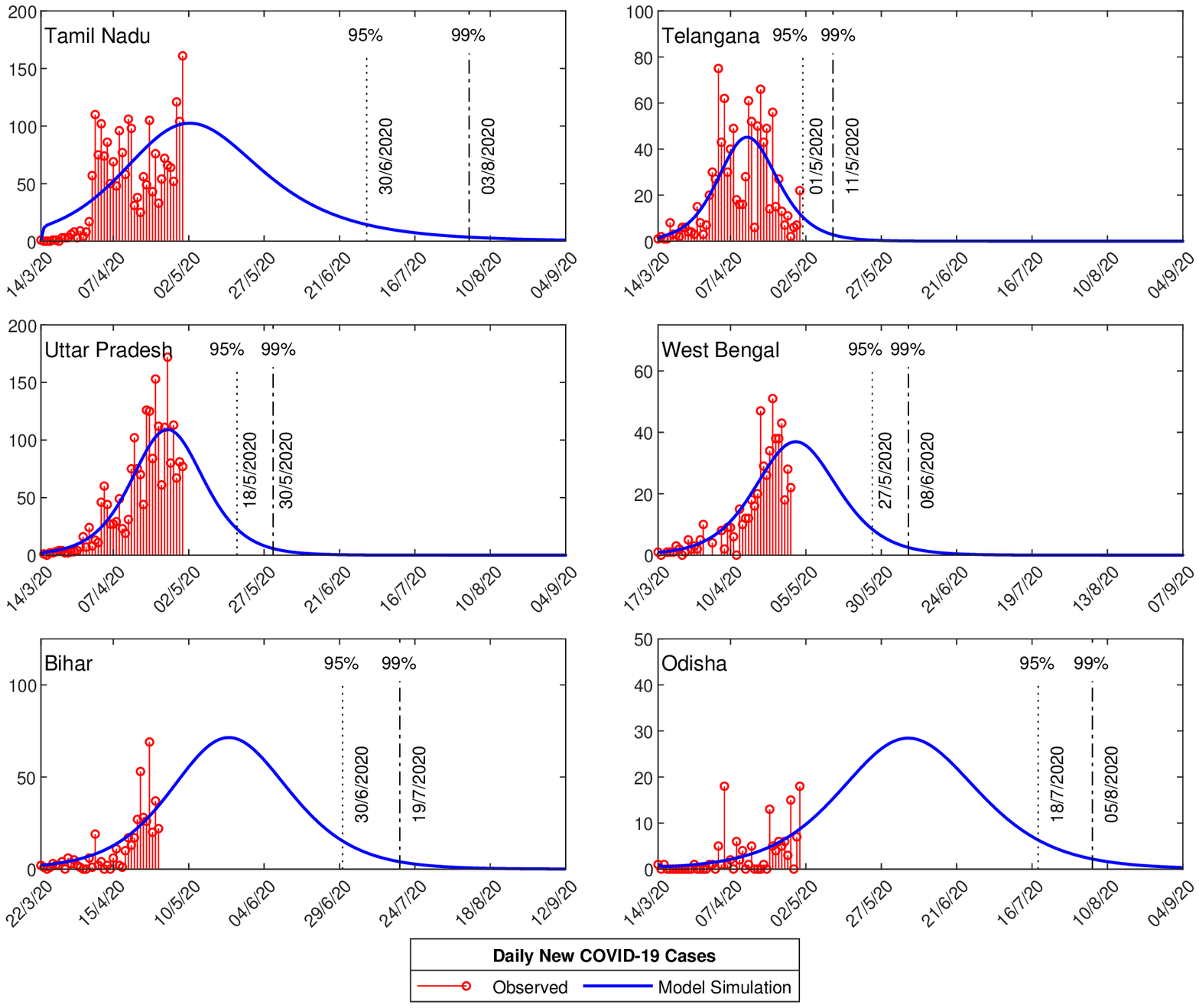}
\caption{Model-Based Data-Driven prediction of SARS-CoV-2 life cycle. Model predictions based on the estimated data for the five provinces of India, namely Tamil Nadu, Telangana, Uttar Pradesh, West Bengal, Bihar, and Odisha. The blue curve represents the model prediction for the daily new confirmed cases of coronavirus diseases while the red dot histogram represents the actual cases. The turning and ending dates of COVID-19 in India and five provinces are displayed in the figure. The estimated parameter values are specified in the Table \ref{parvalEst}.}
\label{f_Predict3}
\end{figure}

Starting from the date of first COVID-19 case reported, we have simulated the $SARII_qS_q$ model for 260 days for each states and for whole India to study the dynamics of the SARS-CoV-2 diseases. The simulated daily  new COVID-19 cases are plotted for 175 days and for 200 days only for Kerala and for whole India in Figures \ref{f_Predict1}, \ref{f_Predict2} and \ref{f_Predict3}. The end date of the pandemic is the last date of COVID-19 case report. However, to take administrative decisions it is more essential to know the proportions of total infected populations rather estimating the end date of the pandemic. We have estimated the dates to reach 95\% and 99\% of the expected total infected COVID-19 cases in India and seventeen states of India. The dates are estimated by computing the total area under the simulated curve and finding the days upto which 95\% or 99\% area are enclosed. The model simulated forecasting with 95\% and 99\% end dates are shown in the Figures \ref{f_Predict1}, \ref{f_Predict2} and \ref{f_Predict3}. The blue curve represents the model forecast for the daily new confirmed cases of coronavirus diseases while the red dot histogram represents the observed cases. The dotted line represents the 95\% end date and dash-dot line represents the 99\% end date.

 The 95\% end date of India is 26 June, 2020 and 99\% end date of India is 27 July, 2020. The 95\% and 99\% end dates are different for different states of India. We have observed that the 95\% end date of states are on or before 26 June 2020 except for the states Andhra Pradesh, Kerala, Punjab, Tamil Nadu, Bihar and Odisha. Also, the 99\% end of thirteen states are on or before 26 July, 2020 but the 99\% end date of Andhra Pradesh, Punjab, Tamil Nadu, and Odisha are 30 July, 1 August, 3 August and 5 August, respectively. That is the 99\% end days deviates 10 days from the 99\% end day of India.

\begin{table}[h]
\caption{\emph{Parameter values estimated from the observed daily new cases of COVID-19 in India and for 17 different provinces of India. Six sensitive parameters $\beta_s$, $\rho_s$, $\varepsilon_s$, $\gamma_a$, $\xi_a$ and $\xi_i$ are estimated among nine system parameters. The parameters are estimated by fitting time series solution of the $SARII_qS_q$ model with the observed daily new cases  of COVID-19 using the least square technique. Also, we compute the basic reproduction number $R_0$ for 17 different provinces of India and the overall India.}}
\begin{center}
 \begin{tabular}{l l l l l l l l}
\hline
 India \& Provinces & $\beta_s$  & $\rho_s$  & $\varepsilon_s$   & $\gamma_a$ & $\xi_a$ & $\xi_i$ & $R_0$ \\
 \hline
India            & 0.8799 & 0.3199  & 14.8300  & 0.01679 & 0.7100 & 0.0286 & 2.0490 \\
Andhra Pradesh   & 1.0228 & 0.1000  & 5.6000   & 0.1000 & 0.4679 & 0.6840 & 1.2014 \\
Delhi            & 0.3100 & 0.3360  & 9.6600   & 0.2840 & 0.6100 & 0.4034 & 1.3301 \\
Gujarat          & 0.5748 & 0.00099 & 9.1140   & 0.1220 & 0.3240 & 0.9276 & 1.4342 \\
Haryana          & 0.6699 & 0.2828  & 13.1220  & 0.1879 & 0.9000 & 0.8948 & 1.1269 \\
Jammu \& Kashmir & 0.5569 & 0.4600  & 12.7564  & 0.1786 & 0.3968 & 0.8520 & 1.2894 \\
Karnataka        & 0.6520 & 0.7360  & 13.3399  & 0.4499 & 0.6740 & 0.7399 & 1.1328 \\
Kerala           & 1.3020 & 0.0090  & 12.7999  & 0.0090 & 0.4060 & 0.2080 & 1.2814 \\
Madhya Pradesh   & 1.0620 & 0.86599 & 8.7600   & 0.2740 & 0.5600 & 0.1780 & 1.6415 \\
Maharashtra      & 1.1400 & 0.1599  & 10.5999  & 0.0099 & 0.1660 & 0.1599 & 2.4690 \\
Punjab           & 1.6099 & 0.1000  & 5.3200   & 0.0208 & 0.6399 & 0.1000 & 1.4149 \\
Rajasthan        & 1.5299 & 0.3100  & 6.7999   & 0.0860 & 0.5020 & 0.7500 & 1.2779 \\
Tamil Nadu       & 0.5499 & 0.8619  & 11.7960  & 0.8660 & 0.6959 & 0.3340 & 1.2248 \\
Telangana        & 0.9160 & 0.7479  & 15.6400  & 0.2000 & 0.5860 & 0.5340 & 1.5177 \\
Uttar Pradesh    & 1.3860 & 0.0208  & 9.0200   & 0.0882 & 0.8700 &  0.7940 & 1.3019 \\
West Bengal      & 1.2180 & 0.4060  & 14.5139  & 0.0202 & 0.2800 & 0.3188 & 1.8120 \\
Bihar            & 1.4100 & 0.5500  & 4.5169   & 0.0900 & 0.0990 & 0.8061 & 1.5552 \\
Odisha           & 0.9700 & 0.0660  & 11.0700  & 0.0380 & 0.3000 & 0.8000 & 1.2938 \\
\hline
\end{tabular}
\end{center}
\label{parvalEst}
\end{table}

\begin{table}[h]
\caption{\emph{Initial population size used in numerical simulations for India and 17 different provinces of India.}}
\begin{center}
 \begin{tabular}{l l l l l l l}
\hline
 India \& Provinces & $S(0)$  & $S_q(0)$  & $A(0)$   & $I(0)$ & $I_q(0)$ & $R(0)$ \\
 \hline
India  & 902654 & 10785 & 114 & 1 & 0 & 0\\
Andhra Pradesh & 94558 & 950 & 95 & 1 & 0 & 0\\
Delhi & 58210 & 505 & 16 & 1 & 0 & 0\\
Gujarat & 68774 & 715 & 76 & 2 & 0 & 0\\
Haryana & 30191 & 351 & 15 & 1 & 0 & 0\\
Jammu \& Kashmir & 21695 & 750 & 12 & 2 & 0 & 0\\
Karnataka & 60156 & 900 & 12 & 2 & 0 & 0\\
Kerala & 32217 & 2151 & 37 & 1 & 0 & 0\\
Madhya Pradesh & 41712 & 978 & 21 & 4 & 0 & 0\\
Maharashtra & 135694 & 1000 & 140 & 18 & 0 & 0\\
Punjab & 21205 & 200 & 110 & 1 & 0 & 0\\
Rajasthan & 103704 & 1100 & 115 & 3 & 0 & 0\\
Tamil Nadu & 119748 & 1197 & 28 & 1 & 0 & 0\\
Telangana & 19278 & 201 & 10 & 1 & 0 & 0\\
Uttar Pradesh & 78013 & 708 & 39 & 1 & 0 &  0\\
West Bengal & 16525 & 575 & 18 & 1 & 0 & 0\\
Bihar & 22672 & 275 & 22 & 2 & 0 & 0\\
Odisha & 34133 & 301 & 5 & 1 & 0 & 0\\
\hline
\end{tabular}
\end{center}
\label{parvalIC}
\end{table}

\begin{table}[h]
\caption{\emph{Sensitivity indices for the basic reproduction number $R_0$ to the parameters for the COVID-19 $SARII_qS_q$ model for 17 different provinces of India and overall India, assessed at the baseline parameter values listed in the Table \ref{parval} and Table \ref{parvalEst}.}}
\begin{center}
 \begin{tabular}{l l l l l l l l}
 \hline
  & & & Sensitivity & Index & & & \\
\hline
 India \& Provinces & $\beta_s$  & $\rho_s$  & $\epsilon_s$   & $\gamma_a$ & $\gamma_{i}$ & $\xi_a$ & $\xi_i$  \\
 \hline
India            & 1.0000 & -0.470372   & 1.0000 & 0.976898 & -0.714314 & -0.976898 & -0.285686 \\
Andhra Pradesh   & 1.0000 & -0.11111    & 1.0000 & 0.823913 & -0.094651 & -0.823913 & -0.905349 \\
Delhi            & 1.0000 & -0.506024   & 1.0000 & 0.682327 & -0.150576 & -0.682327 & -0.849424 \\
Gujarat          & 1.0000 & -0.00099098 & 1.0000 & 0.726457 & -0.0715737 & -0.726457 & -0.928426 \\
Haryana          & 1.0000 & -0.394311   & 1.0000 & 0.827282 & -0.074003 & -0.827282 & -0.925997 \\
Jammu \& Kashmir & 1.0000 & -0.851852   & 1.0000 & 0.689607 & -0.0774328 & -0.689607 & -0.922567 \\
Karnataka        & 1.0000 & -2.78788    & 1.0000 & 0.599697 & -0.0881305 & -0.599697 & -0.911869 \\
Kerala           & 1.0000 & -0.0090817  & 1.0000 & 0.978313 & -0.255841 & -0.978313 & -0.744159 \\
Madhya Pradesh   & 1.0000 & -6.46213    & 1.0000 & 0.671463 & -0.286602 & -0.671463 & -0.713398 \\
Maharashtra      & 1.0000 & -0.190334   & 1.0000 & 0.943718 & -0.309019 & -0.943718 & -0.690981 \\
Punjab           & 1.0000 & -0.11111    & 1.0000 & 0.968518 & -0.416944 & -0.968518 & -0.583056 \\
Rajasthan        & 1.0000 & -0.449275   & 1.0000 & 0.853741 & -0.087047 & -0.853741 & -0.912953 \\
Tamil Nadu       & 1.0000 & -6.24113    & 1.0000 & 0.445547 & -0.176346 & -0.445547 & -0.823654 \\
Telangana        & 1.0000 & -2.96668    & 1.0000 & 0.745547 & -0.118099 & -0.745547 & -0.881901 \\
Uttar Pradesh    & 1.0000 & -0.0212418  & 1.0000 & 0.907952 & -0.0826218 &  -0.907952 & -0.917378 \\
West Bengal      & 1.0000 & -0.683502   & 1.0000 & 0.932712 & -0.183213 & -0.932712 & -0.816787 \\
Bihar            & 1.0000 & -1.2222     & 1.0000 & 0.52381  & -0.0814827 & -0.52381 & -0.918517 \\
Odisha           & 1.0000 & -0.0706638  & 1.0000 & 0.887574 & -0.082053 & -0.887574 & -0.917947 \\
\hline
\end{tabular}
\end{center}
\label{parvalSI}
\end{table}

\begin{table}[h]
\caption{\emph{The values of the errors $E_{MAE}$ and $E_{RMSE}$ to estimate the accuracy of the $SARII_qS_q$ model (\ref{stateeq}) simulation.}}
\begin{center}
 \begin{tabular}{l| l l l l l l}
\hline
 Provinces & India  & Andhra Pradesh  & Delhi   & Gujarat & Haryana & Jammu \& Kashmir \\
$E_{MAE}$  & 114.0705 & 12.2267 & 35.6525 & 27.0626 & 5.0762 & 5.6382\\
$E_{RMSE}$ & 187.4646 & 15.4463 & 54.9932 & 42.3328 & 7.4422 & 7.4386\\
\hline
Provinces & Karnataka & Kerala & Madhya Pradesh & Maharashtra & Punjab & Rajasthan\\
$E_{MAE}$ & 5.5614 & 4.3022 & 30.8349 & 53.2686 & 7.3466  & 21.5003\\
$E_{RMSE}$& 7.7961 & 6.6772 & 44.4765 & 86.4555 & 13.7769  & 27.0582\\
\hline
Provinces & Tamil Nadu & Telangana & Uttar Pradesh & West Bengal & Bihar & Odisha \\
$E_{MAE}$ & 26.5291  & 10.8234 & 15.5550 & 5.2633  & 6.4716 & 2.3084\\
$E_{RMSE}$& 31.0178  & 15.1291 & 23.0103 & 8.6093 & 10.0497 & 3.7892\\
\hline
\end{tabular}
\end{center}
\label{Table:error}
\end{table}

\begin{figure}
\centering
\includegraphics[width=15cm]{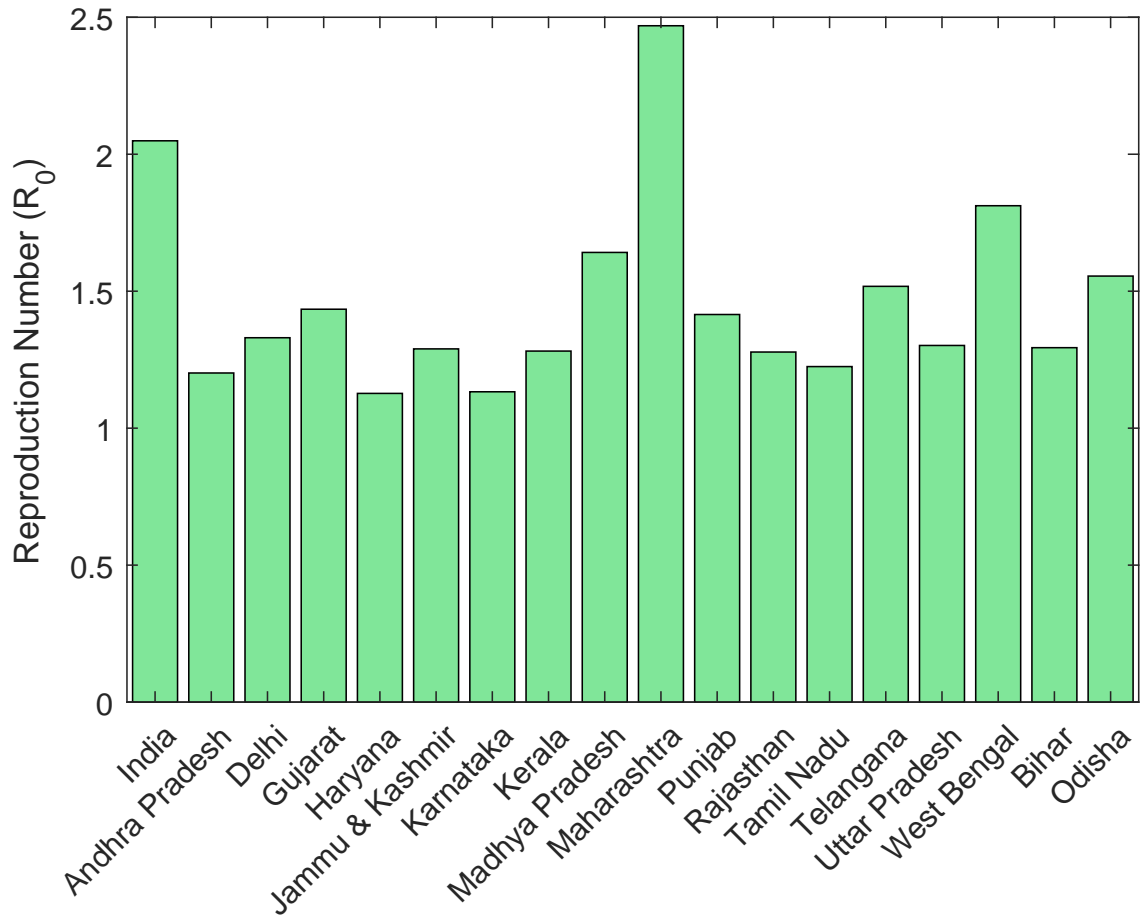}
\caption{Basic reproduction number $R_0$. The bar-diagram represents the computation of basic reproduction number $R_0$ from the estimated parameter values in the Table \ref{parvalEst} for the Republic of India and 17 different provinces of India. }
\label{f_R0}
\end{figure}

\begin{figure}
\centering
\includegraphics[width=15cm]{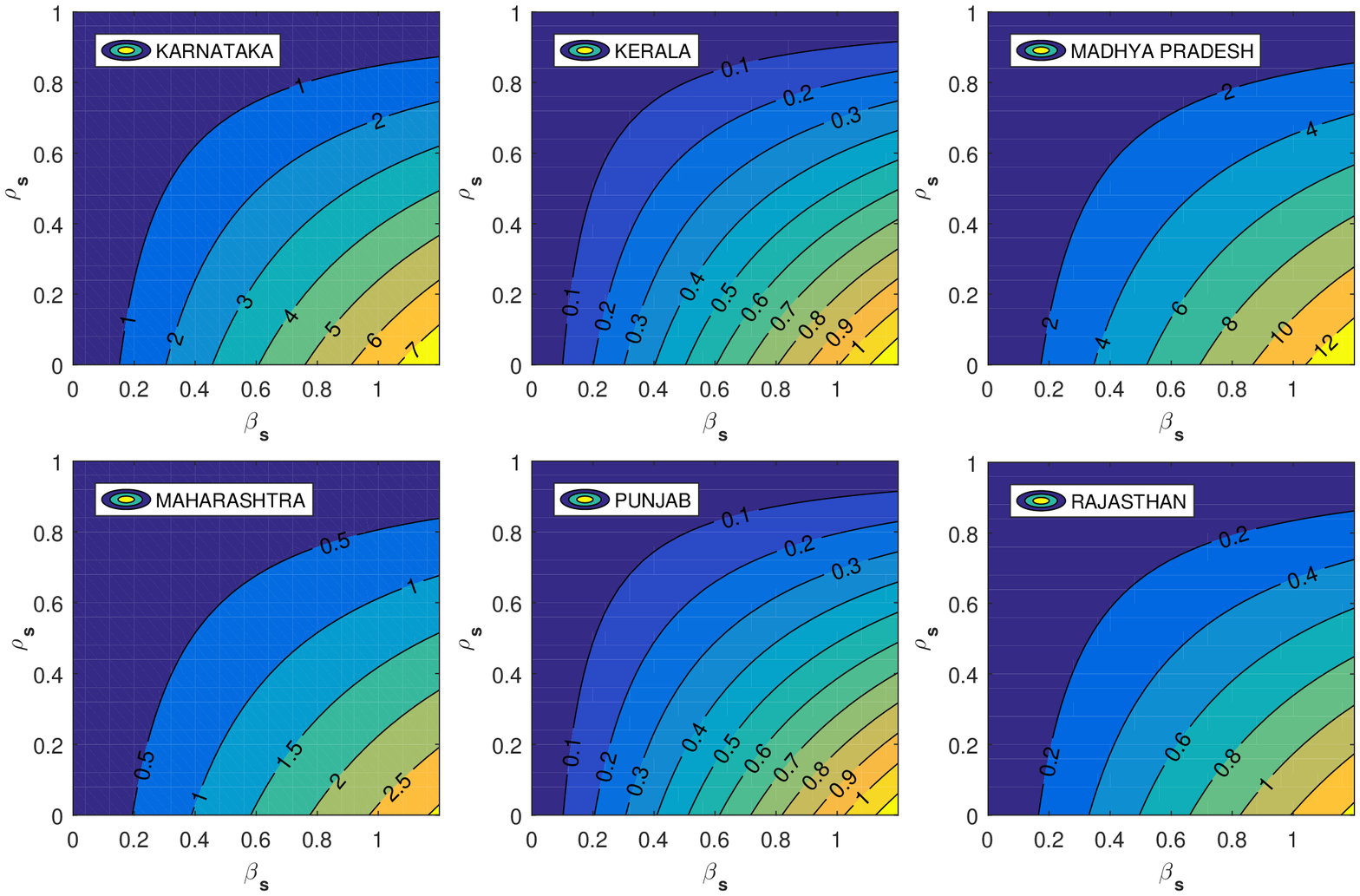}
\caption{Contour plots of basic reproduction number $R_0$. Contour plots of $R_0$ for six different provinces of India, with respect to the probability of disease transmission rate $\beta_s$ and the quarantined rate $\rho_s$ of susceptible individuals. All parameter values other than $\beta_s$ and $\rho_s$ are listed in the Table \ref{parvalEst}. The contour plots demonstrates that higher disease transmission probability of the COVID-19 virus will remarkably increase the reproduction number $R_0$.  }
\label{f_cp2}
\end{figure}

\begin{figure}
\centering
\includegraphics[width=15cm]{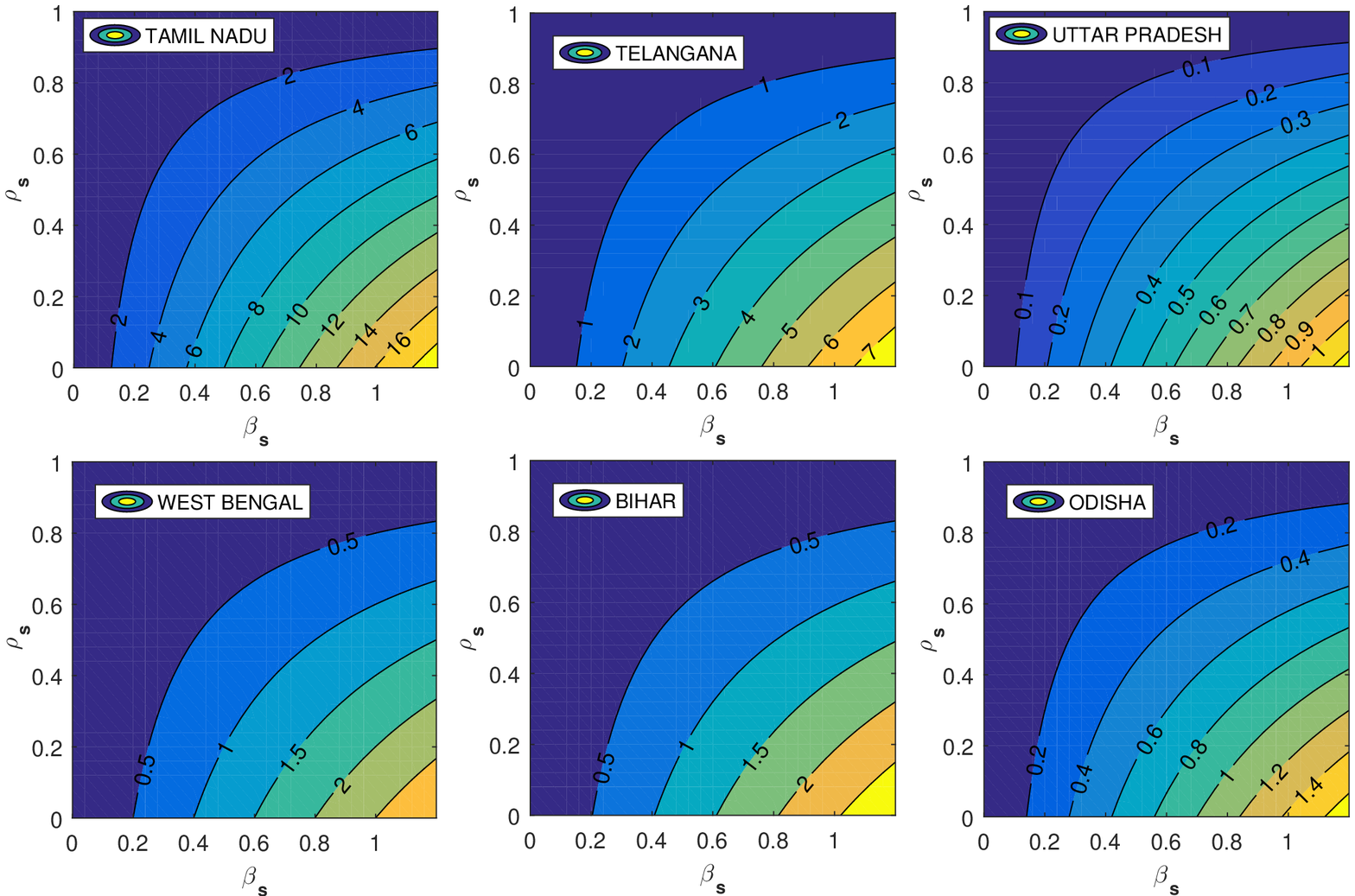}
\caption{Contour plots of basic reproduction number $R_0$. Contour plots of $R_0$ for six different provinces of India, with respect to the probability of disease transmission rate $\beta_s$ and the quarantined rate $\rho_s$ of susceptible individuals. All parameter values other than $\beta_s$ and $\rho_s$ are listed in the Table \ref{parvalEst}. The contour plots demonstrates that higher disease transmission probability of the COVID-19 virus will remarkably increase the reproduction number $R_0$.  }
\label{f_cp3}
\end{figure}

\section{Discussion}

We have investigated a methodology in estimating the crucial epidemiological model parameters as well as the modeling and prediction of the spread of SARS-CoV-2 outbreak in 17 different provinces of India and the overall India by evaluating publicly accessible real data till April 30, 2020.

Our model simulations provide an important tool to evaluate the consequences of possible policies, incorporating social distancing and lockdown. Our model findings, obtained by combining the $SARII_qS_q$ model with the accessible data regarding the coronavirus outbreak in India, indicate that imposing social distancing is crucial, essential and effectual, in line with other details in the literature \cite{Wu20}. If the strong lockdown can be imposed earlier, the effective results can be achieved. We accept these indications can be handy to control the outbreak in India, together with the countries that are still in the early phases of epidemic.

The future is always unpredictable and we must keep this in mind when we read any kind of predictions. No one forecasted the outbreak of COVID-19 in India and worldwide, however Bill Gates warned about the potential harm of a worldwide infectious disease in a speech in 2015. Based on the real data on COVID-19 pandemic in India till April 30, 2020 we predicted our $SARII_qS_q$ model. The total forecasted infected individual size is the total area under the curve. Our model forecasts give the following two estimates of end dates: (i) the date to reach 95\% of expected total number of cases and (ii) the date to reach 99\% of expected total number of cases. At the end, we want to conclude that the number of infected individuals may increase in India and the 17 provinces of India for the next few months if human-to-human transmission and personal preventive measure resume with the existing rates as the testing status are very poor in India. It is worthy to mention that the researchers around the world is working for effective vaccine and/or therapeutics against coronavirus or SARS-CoV-2 diseases and the existence of such pharmaceutical interventions will remarkably change the results.

In the modern and developed world, data and detailed information regarding the COVID-19 or SARS-CoV-2 and the evolution of epidemic become accessible at an unparalleled pace.  Howbeit, important questions still remain undetermined and precise answers for forecasting the transmission dynamics of the epidemic simply cannot be acquired at this stage. We emphasize the uncertainty of accessible authentic data, specially concerning to the accurate baseline number of infected individuals, which may guide to the equivocal outcomes and inappropriate predictions by orders of size, as also identified by the other researches \cite{Battegay20}. We hope that our predictions will be handy for Govt. and different companies as well as the people towards making resolutions and considering the suitable actions that contain the spreading of the coronavirus to the possible stage.



\end{document}